\documentclass[iop]{emulateapj}
\shorttitle{F/F' RV Variability measured with SORCE}
\shortauthors{Marchwinski, R. C., Mahadevan, S., Robertson, P., Harder, J., Ramsey, L.}
\begin{document}
\title{Towards Understanding Stellar Radial Velocity Jitter as a Function of Wavelength: The Sun as a Proxy}
\author{Robert C. Marchwinski, Suvrath Mahadevan, Paul Robertson, Lawrence Ramsey}
\affil{Department of Astronomy \& Astrophysics, The Pennsylvania State University, 525 Davey Lab, University Park, PA 16802}
\affil{Center for Exoplanets and Habitable Worlds, The Pennsylvania State University, 525 Davey Lab, University Park, PA 16802}
\and
\author{Jerald Harder}
\affil{Laboratory for Atmospheric and Space Physics, University of Colorado Boulder, 1234 Innovation Drive, Boulder, CO 80803}
\slugcomment{Submitted to the Astrophysical Journal}
\received{8 Aug 2014}
\accepted{27 Oct 2014}
\begin{abstract}
Using solar spectral irradiance measurements from the SORCE spacecraft and the  F/F' technique, we have estimated the radial velocity (RV) scatter induced on the Sun by stellar activity as a function of wavelength.  Our goal was to evaluate the potential advantages of using new near-infrared (NIR) spectrographs to search for low-mass planets around {\bf bright} F, G, and K stars by beating down activity effects.  Unlike M dwarfs, which have higher fluxes and therefore greater RV information content in the NIR, solar-type stars are brightest at visible wavelengths, and, based solely on information content, are better suited to traditional optical RV surveys. However, we find that the F/F' estimated RV noise induced by stellar activity is diminished by up to a factor of 4 in the NIR versus the visible. Observations with the upcoming future generation of NIR instruments can be a valuable addition to the search for low-mass planets around bright FGK stars in reducing the amount of stellar noise affecting RV measurements.
\end{abstract}
\section{Introduction \& Background}
The first detections of exoplanets are now 2 decades old \citep{CWY88,LATH89,WF92,MQ95} and over 700 confirmed and $\sim$3000 candidate exoplanets \citep{WFM11} are now known. This number is only increasing due to ongoing and planned radial velocity \citep[RV;][]{HARPS,CWY88}, transit \citep{KEP10,KTJS03}, microlensing \citep{BUJ04}, and direct imaging \citep{KGC08} surveys. The search is now focused on detecting low mass planets, with great interest in those that orbit in a star's habitable zone \citep[HZ;][]{HZ93,SELSHZ,KOPP13}. The fundamental limitation on the detection of these low mass exoplanets is no longer just instrumental, but also astrophysical \citep{HATZES13}.

Stellar analogs of solar activity features such as faculae, starspots, stellar flares, and stellar cycles induce time dependent variations on the observed brightness and spectra of stars. These can be large enough to mask or imitate planetary companions \citep{QUEL01,IF10,DUMUSQ12,HATZES13,ROBERT13} with the largest impact on the detection of low mass planets. Stellar models, long term observations, and higher resolutions can help address these effects \citep{HAT99}, but often at high observational cost. 

A well known example is the proposed Earth-mass planet orbiting $\alpha$ Cen B in a 3-day orbit \citep{DUMUSQ12}. \citet{HATZES13} showed that the significance of the signal depends on the method used to treat the activity-induced RV from the star, with some epochs exhibiting an activity-induced RV amplitude of 7 m s$^{-1}$. Stellar noise has also led to controversy regarding several of the low mass planet candidates orbiting the M dwarf Gl 667C \citep{AS13,DELF13} with activity related signals discussed by \citet{PAULG}.

The system of most controversy has been Gl 581 and its multiple, proposed, HZ super-Earths. Recently, these have been shown to actually be stellar activity signals \citep{PAULK}. This result not only shows how stellar activity can appear to be planetary signals, but correcting for these stellar signals boosts the signal of the real planets. Identifying how to minimize astrophysical systematics, especially with regard to stellar activity, is thus extremely important to confirm the small exoplanets, as also shown recently by \citet{HAYW14} for the COROT-7 system.

Over the past few centuries studying our Sun has helped us to understand stellar activity and stellar processes. Initial observations involved counting sunspots, which was later followed by the identification of large flares, and the space age bringing details about coronal mass ejections (CMEs). \citet{PARK55} tied most of the variable characteristics of the Sun to the solar magnetic field through the $\alpha\Omega$ dynamo \citep{STIX76}. The $\alpha\Omega$ dynamo is based on magnetic field lines being stretched, wound, and twisted by the Sun's rotation, especially differential rotation between the Sun's convective and radiative layers. Another small scale variation is contributed by granulation on the surface due to the convective nature of the underlying layer \citep{BRAY84}. Both of these effects manifest as photometric variability for the Sun and for other stars \citep{FB13}. 

 New near-infrared (NIR) precision radial velocity spectrographs are slated to come online this decade. With the advent of better technology, specifically in calibration, NIR instruments are now being built to enable RV precision approaching that currently obtained in the optical. In addition to leveraging the greater NIR flux of late K and M stars, these spectrographs are expected to benefit from a reduction in activity-induced RV in the NIR. This work explores the question: can the new NIR spectrographs, such as the Habitable Zone Planet Finder \citep[HPF;][]{HPF10}, the Calo Alto high-Resolution search for M dwarfs with Exo-earths with Near-infrared and optical Echelle Spectrographs \citep[CARMENES;][]{CARM10}, and the infrared spectropolarimeter \citep[SPIRou;][]{SPIR12} also help observe and successfully detect or confirm low mass exoplanets around bright solar type stars by ameliorating the effects of stellar activity? A conclusive answer to this question will be important in the design of the next generation of NIR diffraction limited spectrographs coupled to large telescopes.

Using data provided by the SOlar Radiation and Climate Experiment spacecraft \citep[SORCE;][]{SORCE1}, we have undertaken a study of solar noise as a function of wavelength and time. We estimate RVs from SORCE data using the photometric F/F' technique \citep{AIGRAIN12}. Section 2 briefly discusses our understanding of solar and stellar activity, \S 3 discusses the F/F' Method, and in \S4, the SORCE observations and spacecraft are described. Section 5~discusses the analysis including the implementation of the F/F' method. The results are presented in \S6 and \S7 contains a discussion. In \S8, this work is summarized and future potential work is presented.

\section{Stellar Activity}
The Sun was long considered to have a constant irradiance, i.e. the solar constant, but was also known to be variable through sunspots and solar flares. The advent of space-based radiometers has allowed the small changes in the solar constant to be measured precisely. These changes can be tied to physical features including sunspots, p-mode oscillations, granulation, faculae, plages, and supergranulation \citep{SOLUN13}. It is generally assumed that similar features are responsible for variations observed on other stars as well, and are driven by the magnetic field, $\alpha\Omega$ dynamo, and convection. 

These features manifest on different timescales and in different observable ways. Temporally, p-mode oscillations and granulation show variations on the shortest timescales (minutes to hours). Granulation, sunspots, faculae, and plages are responsible for the variation up to a few days, with sunspots driving variability on timescales similar to the rotation period \citep{SOLUN13}. These effects are known to be wavelength dependent. Sunspots change the observed brightness highly in the visible but less in the infrared \citep{REI10} while granulation may affect all of the spectrum. 

These features also affect observations of the surface kinematics \citep{LEI62}. Granulation is an upwelling driven by the motion of the solar material \citep{LEI62}, while sunspots, faculae and other magnetically driven phenomena can add or subtract flux with different Doppler velocities. The decrease or addition of velocity shifted emission is observed in other stars and alters measurements of the absolute RVs \citep{SD97,HAT02}. Any activity-induced shift in RV; random, coherent, or periodic will impact the detection of planets around other stars. Simulations of the impact of stellar activity on RVs have been conducted by \citet{LM10a}, \citet{LM10b}, \citet{MEU10}, and \citet{CEG14}. We seek in this work to use actual spectro-photometric observations of the Sun to explore the RV variability as a function of wavelength.

\section{F/F' Method}
Despite the plethora of observations of the Sun, obtaining an integrated RV measurement is difficult. The resolved nature of the Sun means one must either integrate the flux over the Sun's solid angle or observe reflected light from other objects. The first was demonstrated using the Solar Dynamics Observatory \citep[SDO;][]{MEU10,SDO1} and showed agreement with simulations to within 30\%. Preliminary results from the second method are detailed by \citet{RVAST} using the High-Accuracy Radial velocity Planetary Searcher spectrograph \citep[HARPS;][]{HARPS} and reflected light from the asteroid Vesta. This method introduces additional factors that must be accounted for such as the asteroid's properties, its rotation, and its velocity. More importantly, obtaining the measurements needed for a long-term RV characterization of the Sun is observationally expensive requiring many nights of observing to acquire sufficient data.

In this work we estimate solar RV variability as a function of wavelength using the F/F' method developed by \citet{AIGRAIN12}. This technique relies on a simplified spot model that converts observed stellar fluxes into estimated RV variations. Starspots, being cooler than their surroundings, emit less light. Because these spots are spatially distributed, the lower emission comes from specific regions that would normally have a Doppler shift due to rotation and convection. The spots suppress these regions's contributions to the integrated flux, shifting the measured RV. This effect can cause scatter or coherent signals that can suppress or masquerade as the signals of planets.

The F/F' method attempts to replicate the RV variability that would be observed if the Sun was an unresolved source. The governing formulae are:

\begin{equation}
\Delta RV(t) = \Delta RV_c(t) + \Delta RV_{rot}(t) 
\label{eq:one}
\end{equation}
\begin{equation}
\Delta RV(t) = \frac{\delta V_c \kappa}{f}\left[1-\frac{\Psi(t)}{\Psi_0}\right]^2 + \frac{\dot{\Psi(t)}}{\Psi_0}\left[1-\frac{\Psi(t)}{\Psi_0}\right]\frac{R_\star}{f}
\label{eq:two}
\end{equation}

Equation \ref{eq:one} shows that the total RV variation ($\Delta RV(t)$) estimated by the F/F' method is the sum of the estimated convective component ($\Delta RV_c(t)$) and the estimated rotational component ($\Delta RV_{rot}(t)$). The convective component comes from the suppression of the convective blueshift in spot regions while the rotational component is the RV signature of spots reducing light from different regions with different rotational velocities. These terms are then broken down in Equation \ref{eq:two} into the components that can be measured from the photometry or modeled. $\delta V_c$ is the difference between the convective blueshift in the spot region and the normal photosphere, $\kappa$ is the ratio of the magnetized area to the spot surface, $\Psi_0$ is the maximum measured flux, $\Psi(t)$ is the measured flux, $\dot{\Psi(t)}$ is the derivative of the flux, $f = (\Psi_0 - \Psi_{min})/\Psi_0$, where $\Psi_{min}$ is the minimum observed flux, and $R_\star$ is the radius of the Sun.

While the F/F' technique can be used to estimate an RV shift, it is worth emphasizing that it yields only an incomplete (albeit powerful) account of activity related RV noise. For example, it does not consider the photometric effect of faculae (which can lead to the suppression of the convective blueshift). \citet{HAYW14} invoked an additional activity function to supplement the F/F' technique in their analysis of the COROT-7 system, though this is only useful if simultaneous precision photometry and RV data are available. \citet{AIGRAIN12} demonstrated that the F/F' technique alone can give results equivalent to those of long term, conventional RV measurements. Data from the Microvariability and Oscillations of Stars mission \citep[MOST;][]{MOST03} and observations from the SOPHIE spectrograph \citep{PER08} showed good agreement between the spectral and photometric RVs for HD 189733 \citep{AIGRAIN12}. In addition, the F/F' method showed good agreement when comparing synthetic solar photometry and RV datasets. We use the F/F' technique to calculate solar RV variability derived from space-based spectrophotometric observations from the SORCE spacecraft. While this ignores sources of RV variability that the F/F' technique does not account for, we believe it is suitable for our purpose which is to probe the RV variability as a function of wavelength.

\section{SORCE}
SORCE is a NASA sponsored spacecraft mission launched in 2003 as part of NASA's Earth Observing System \citep{SORCE1}. The main mission of SORCE is to measure the incoming solar radiation, its variations, and its properties from the x-ray and ultraviolet wavelengths through the near-infrared

The SORCE spacecraft itself consists of four instruments: the Total Irradiance Monitor \citep[TIM;][]{KOPP05}, the Spectral Irradiance Monitor \citep[SIM;][]{HAR05}, the SOLar STellar Irradiance Comparison Experiment \citep[SOLSTICE;][]{MRW05}, and the XUV Photometer System \citep[XPS;][]{WRV05}. This work uses data from SIM and TIM. Because SORCE was built for a different science goal than we are using it for, the instruments, specifically SIM, must be well understood to derive results accurate for our needs.

\subsection{Total Irradiance Monitor (TIM)}
TIM is responsible for measuring the total solar irradiance (TSI). TIM does this using an electrical substitution radiometer (ESR) with 4 bolometers and an aperture of 0.5 cm$^2$ (total incident flux $\sim$65mW). The ESR uses these 4 bolometers in 2 pairs. Within each pair, one bolometer is exposed to sunlight and allowed to absorb the incident irradiance. The other bolometer is then heated to match the first. The amount of energy needed to bring the bolometers into equilibrium is equal to the amount of solar irradiance absorbed. Using this setup, TIM is able to measure the TSI to an accuracy of parts per million \citep{SORCE1} while maintaining accuracy and measuring (and compensating) for degradation using the two pairs. This is some of the highest precision photometry ever achieved, exceeding that of the \emph{Kepler} mission by a factor of $>$5. For full instrument description see \citet{KOPP05}.

\subsection{SIM: Hardware Description}
SIM is the primary instrument of interest for this study. It observes the SSI over a wavelength span of 200 to 2700 nm \citep{HAR05}. Its design is a dual Fery prism spectrometer. To observe over this wavelength span, SIM uses 5 different detectors and observes an integrated irradiance of $\sim$1320 W m$^{-2}$.\footnote{See \S \ref{sec:timsim} for discussion of missing irradiance} Solar radiation incident on the instrument can take one of three paths through the instrument. These paths allow the spectral irradiance to be measured (spectrometer path), the instrument to be wavelength calibrated (wavelength control path), or the instrument's degradation to be measured \citep[prism transmission path;][]{HAR05}.

The primary detector, used for calibration and large wavelength measurements, is an ESR which uses two independent bolometers and initially places them in temperature equilibrium via heating from resistors. One of the surfaces is then exposed to the radiant energy of the sun. The heating from the resistor for the active bolometer is then reduced until the surface temperatures return to equilibrium. The electrical power reduction is then equal to the radiant energy incident on the bolometer to within the noise limit \citep{HAR05}.  The ESR can measure the irradiance over the total 255-2700 nm range, however, it is not suitable for spectral scanning purposes due to its slow response time. Therefore, the ESR is used more for calibration of the other 4 detectors.

  \begin{deluxetable}{ccc}
\tabletypesize{\footnotesize}
	\centering
	\tablewidth{0pt}
			\tablecolumns{3} 
			\tablecaption{SIM Interruption Events}
	\tablehead{\colhead{Date} &
						\colhead{} &
						 \colhead{Spacecraft/Instrument Event} \\
						 \colhead{UT} & 
						 \colhead{}  &
						 \colhead{}  }
		\startdata 
			 2003/04/19.1 & & Onboard computer (OBC) reset  \\
			 2003/08/13.5 & & Overheat anomaly  \\
			 2003/10/28.6 & & Cold soak experiment \\
			 2004/04/18.0 & & Prism rotation encoder corrected \\
			 2007/05/14.9 & & OBC reset \\
			 2009/01/05.0 & & OBC reset \\
			 2009/10/14.5 & & Safehold, Battery under-voltage \\
			 2010/09/27.0 & & OBC reset \\
			 2010/11/01.0 & & OBC reset \\
			 2010/12/27.0 & & OBC reset \\
			 2011/01/29.0 & & OBC reset \\
			 2011/05/16.0 & & CPV10 failure (battery failure) \enddata
			 \label{tab:events}	 
\end{deluxetable}

The other 4 detectors used by SIM are 3 silicon photodiodes and 1 InGaAs photodiode. Two of the silicon photodiodes are designed with n-p silicon geometry along with a nitride passivated SiO$_2$ layer to make them sensitive and stable in the UV. These two detectors, UV and Vis1, cover wavelengths of 200-308nm and 310-1000nm respectively. The final silicon photodiode (Vis2) has p-n silicon geometry and has a wavelength coverage of 360-1000nm. Finally, SIM has an InGaAs photodiode used for IR irradiance measurements. This detector has a wavelength coverage of 944-1655nm. These 5 detectors allow SIM to measure the SSI with high accuracy over 97\% of the solar spectral energy distribution.

The resolution of SIM changes over the wavelength range covered. From 200 to 400 nm the FWHM of the resolution elements is less than 5nm. The resolution increases up to a peak of around 20-30nm near 1200nm, before slowly decreasing.

The SIM mission began in 2003 and lasted until 2011 at which time a power cycling mode was initiated due to a battery failure. Therefore, the SIM database provides an almost continuous dataset of daily solar spectral irradiance measurements for nearly 8 years with the interruptions in the observations listed in Table \ref{tab:events}. 

 \begin{figure}
\begin{center}
\includegraphics[trim=1cm 3.4cm 2cm 3.8cm, clip=true,width=0.99\columnwidth]{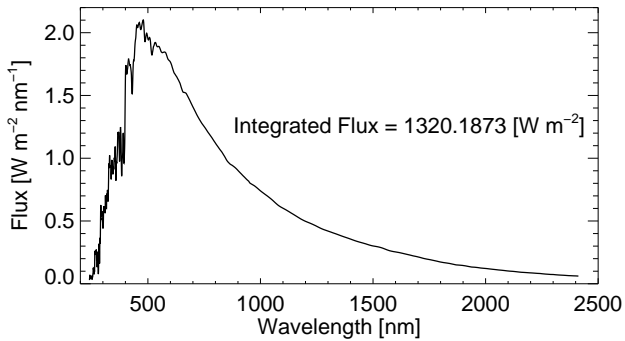}
\caption{SIM data for 2006 January 29 shown as flux vs. wavelength. The integrated flux value is also shown.}
\label{fig:sim}
 \end{center} 
 \end{figure}

\subsection{SIM: Data Artifacts}
\label{sec:SIM}
Due to SIM's design, a number of systematic effects have to be taken into account when using the data for our needs. Specifically, in the 800-1000nm region the quantum efficiency of the photodiodes is strongly dependent on their temperature due to the way they  are biased. This introduces an increased systematic uncertainty that cannot fully be accounted for and the data in this region have been excluded from our analysis. Our analysis showed that this effect is still present down to 760nm. Additionally, the ESR is the only detector available for measurements past 1600nm. Wavelengths longer than 1600nm have not been included in our analysis due to the ESR's larger uncertainty.

\subsection{TIM and SIM}
\label{sec:timsim}
TIM measures TSI every 6 hours in addition to providing a daily value. This allows comparisons to be made between spectral frequency and total irradiance variations (ie.  comparison of the SIM and TIM measurements). Integrating the SIM measurements yields an average measured irradiance of 1319.1 W m$^{-2}$ (97\% of the measured TSI) with a standard deviation of 0.63 W m$^{-2}$ over the entire mission. The SIM flux values over the entire wavelength coverage for 2006 January 29 are shown as an example in Figure \ref{fig:sim} along with the integrated flux for that day. The average TSI value is 1360.7 W m$^{-2}$ with a standard deviation of 0.3 W m$^{-2}$. The missing 40 W m$^{-2}$ of energy from SIM is distributed with $\sim$14 W m$^{-2}$ emitted in the FUV and X-rays, and the remainder emitted in the infrared tail.

\section{Analysis}
The SIM and TIM datasets were retrieved from the SORCE website and provided daily measurements of the total and spectral irradiance from 2003 April to 2011 April. These data sets were fairly continuous, with breaks described by the instrument and spacecraft events in Table \ref{tab:events}. Each data set provides the time of observation and the irradiance value (either total or at each wavelength of observation) with the same cadence as the synthetic solar data used by \citet{AIGRAIN12}. The same technique as used by \citet{AIGRAIN12} can therefore be applied to the SORCE data to estimate the RV variations that would be observed due to sunspots and convection.

To estimate the RV shift, the quantities described in Equation \ref{eq:two} need to be calculated or estimated. These quantities estimate the rotational and convective RV variations. The measured flux $\Psi(t)$, the maximum measured flux $\Psi_0$, and $f$ were taken from the observed SIM and TIM irradiance datasets. $\dot{\Psi(t)}$, is the derivative of the irradiance and was found from the SIM and TIM light curves using a simple spline model fit and its derivative. To mitigate transient effects from introducing large changes in the derivative, the light curve was smoothed with a 3-day boxcar. This is the same technique as applied in \citet{AIGRAIN12} where the smoothing width is equal to approximately one-tenth the rotation period. An additional check on the behavior of the derivative was made using the difference between neighboring points. Both results produced similar amplitudes for the derivative. The final quantity that is known is $R_\star$, the radius of the Sun \citep{EMI12}.

The two quantities that are not known precisely are $\kappa$ and $\delta V_c$. These two quantities significantly affect $\Delta RV_c(t)$ and therefore $\Delta RV(t)$ as well. \citet{AIGRAIN12} fit for $\kappa$ and $\delta V_c$ as one quantity with values of $\kappa \delta V_c$ ranging from $\approx10$ to $400$ m s$^{-1}$. For our calculations, we chose $\delta V_c \kappa=$ 400 m s$^{-1}$, which was fitted for the active solar case. This value is important as it increases the contribution of the estimated convective RV scatter to the total RV scatter. This value was adopted for a number of reasons; \citet{AIGRAIN12} fitted for this value using synthetic TSI values thus removing any dependance on our data set and these synthetic values overlapped with our observations so should be accurate for the solar cycle of interest. We do note that the choice of a constant does affect the time-dependent analysis as $\kappa$ does change with the solar cycle and with wavelength. This should only have the effect that the values estimated during the solar quiescent period are to be taken as upper limits.
 
\section{Results}
\begin{figure*}
\begin{center}
\includegraphics[trim=2cm 3.8cm 1cm 3.8cm, clip=true,width=0.99\textwidth]{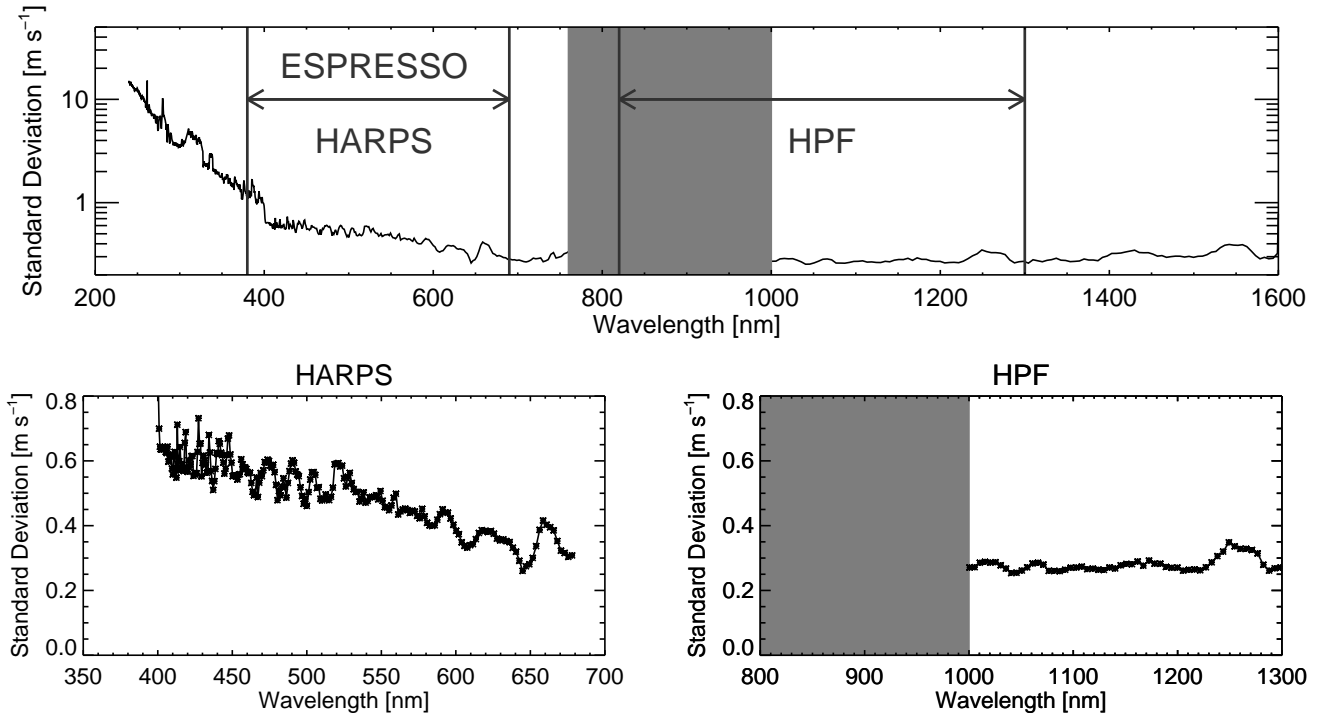}
\caption{\emph{Top:} Semi-log plot of the SSI $\sigma_{\textrm{RV}}$ over the entire mission shown from 200-1600 nm. The grey region covers where detector effects make the data unreliable. Also shown are the wavelength regions covered by HARPS (378-691 nm), the future ESPRESSO (380-686 nm), and the future HPF (820-1400 nm). \emph{Bottom Left:} Zoom in on the HARPS region in non-log space. \emph{Bottom Right:} Zoom in on the HPF covered region in non-log space.}
\label{fig:stdev}
 \end{center} 
 \end{figure*}

The analysis described above resulted in estimated $\Delta$RVs from the TSI and SSI as a function of time, the latter also as a function of wavelength. Figure \ref{fig:stdev} shows the standard deviation of the SSI $\Delta$RVs over the entire mission. The greyed regions are excluded due to additional noise sources from detector effects (See Section \ref{sec:SIM}).

This figure shows that the $\Delta$RVs decrease in variability with increasing wavelength. In the UV higher variability is seen, decreasing throughout the visible, and eventually reaching a floor around a standard deviation of 0.3 m s$^{-1}$. The subplots show the data in the HARPS and HPF bandpasses. The visible wavelengths covered by HARPS still contain significant variation. This variability has an inverse relationship with increasing wavelength. The NIR bandpass shows little variation with wavelength. Additionally, the observed NIR variation has a lower overall magnitude than in the visible. This data suggests that when searching for low mass planet RV signals, the NIR will have a lower stellar RV component. Some caveats are important in making this claim:
\begin{itemize}
\item{The advantage of the NIR spectrographs becomes truly important if they can reach levels of precision comparable, or not significantly worse, than the state of the art in optical RV spectrographs. This statement hold both for instrument limits on RV precision as well as photon \& information content limited RV precision (see Sec 7.4 for a more detailed discussion of the latter).}
\item{Due to the lower information content in the NIR and red for F, G, and K stars as compared to the remainder of the optical \citep{BOUCH01}, NIR instruments will have to be on larger telescopes or be limited to studying activity on the brightest stars to identify the smallest planets. This condition does not apply to the mid to late M dwarfs where NIR instruments are competitive with optical spectrographs from an information content perspective.}
\item{We assume observed stellar activity is magnetic activity and is manifested by spots, faculae, prominences, and flares with driving physics similar to our Sun albeit of varying magnitudes.}
\end{itemize}

 \begin{figure}
\begin{center}
\includegraphics[width=0.45\textwidth, trim = 2cm 3cm 1.5cm 3cm]{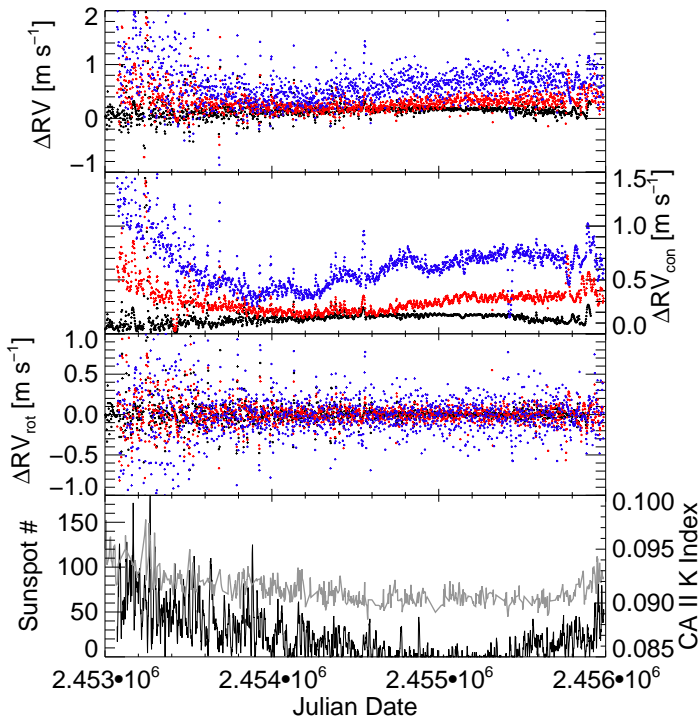}
\caption{\emph{Top Panel:} Estimated $\Delta$RV vs. time from the F/F' method for the HARPS bandpass (blue), HPF bandpass (red), and for the TSI (black). Every other day is plotted for clarity. \emph{Second Panel:} $\Delta$RV contribution from the convective term vs. time. \emph{Third Panel:} $\Delta$RV contribution from the rotational term vs. time. \emph {Bottom Panel:} Number of sunspots observed per day (black, http://www.swpc.noaa.gov/SolarCycle/) and \ion{Ca}{2} K emission index \citep{CALKII}. We draw the reader's attention to the inverse relationship between sunspot number and convective $\Delta$RV.}
\label{fig:comp}
 \end{center} 
 \end{figure}
 \newpage
\section{Discussion}
The wealth of data from the SORCE mission, with the F/F' method, has permitted the estimation of the RV variation that would be observed if we were observing the Sun from a distant planet. The main result from this work is that the RV variability in the NIR is less than that in the optical for the Sun, and by proxy for solar type stars..  We further examine our results, with a focus on time variability, to better understand the relative contributions from the two components used in the F/F' method. 

\subsection{Components used to Estimated RV variability}
The F/F' method consists of two components to the total estimated $\Delta$RV, that from convection and from rotation. The convective term is a blueshift from the upwelling of material seen at the photosphere. The rotation term is due to spot coverage and relies on the decreased emission from spot regions associated with a particular radial component of the rotational velocity. It is the combination of these terms that the F/F' method uses to derive the RV variability. Figure \ref{fig:comp} shows the total $\Delta$RV, the convective $\Delta$RV, and the rotational $\Delta$RV over our entire baseline. 

The top panel of Fig. \ref{fig:comp} shows that the TSI consistently estimates the lowest $\Delta$RV. This is expected as activity mechanisms influence the blue wavelengths more than the integrated output of the Sun. The SIM and TIM data show that the TSI changes by $\sim$1.3 W m$^{-2}$ while the UV (240-400nm) and visible (400-750nm) each change by $\sim$2 W m$^{-2}$, though in opposite directions, over the solar cycle. The UV flux correlates with TSI, decreasing towards solar minimum, while the visible increases during the same time period and with the same amplitude showing evidence for this higher variability. Therefore, the TSI estimated $\Delta$RVs should represent a floor when using the F/F' method. We note that this is different from the case of simultaneous photometry from Kepler or CoRoT as these bands are fairly well matched to the optical RV instruments used to confirm planets. Also seen is that the NIR $\Delta$RVs are lower than the visible $\Delta$RVs consistently over the entire mission, in agreement with Fig. \ref{fig:stdev}. 

For the convective $\Delta$RVs, the same wavelength dependence applies as for the total $\Delta$RV. We draw the reader's attention to the number of sunspots and the \ion{Ca}{2} K emission index \citep{CALKII} plotted in the bottom panel. We can see that, except for the initial values, the number of sunspots has an inverse relationship to the convective $\Delta$RV. 

Finally the third panel shows the rotational $\Delta$RV contribution. This plot shows that the TSI, NIR, and visible appear indistinguishable. We again see a relationship with the number of sunspots. As the number reaches a minimum in the solar cycle, the rotational term shows less scatter approaching zero. This is expected as when there are few to no sunspots, the mechanism that the F/F' method relies on for this contribution is not present.

\subsection{Monthly Data and the Solar Cycle}
\begin{figure*}
\begin{center}
\includegraphics[trim=0cm 3.6cm 1.5cm 4cm, clip=true,width=0.99\textwidth]{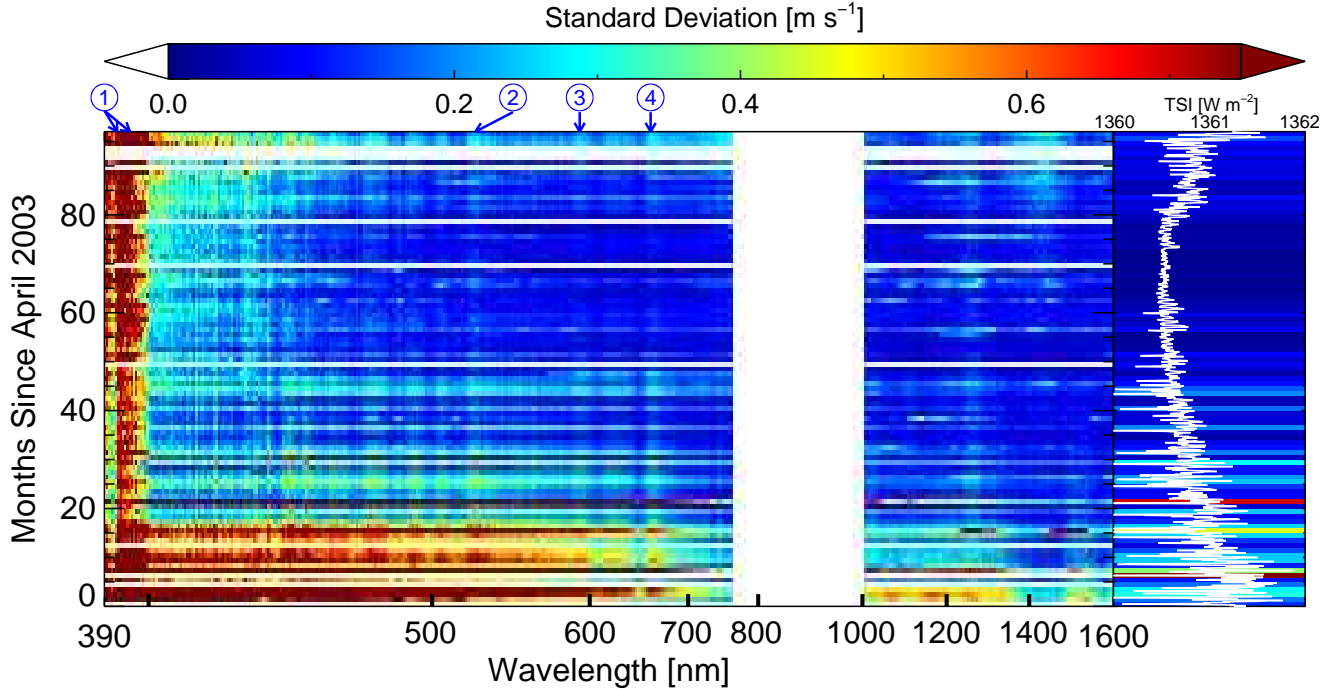}
\caption{Density plot of $\sigma$RV by month. The color represents the value for a given wavelength and month since April 2003. White regions are excluded data including months of spacecraft events. The blue circled numbers correspond to the following lines: 1, \ion{Ca}{2} H \& K; 2, \ion{Mg}{1} triplet; 3, \ion{Na}{1} Doublet; and 4, H$\alpha$. Right plot shows the same $\sigma$RV per month derived for the TSI, with the actual TSI over plotted in white.}
\label{fig:monthly}
 \end{center} 
 \end{figure*}

The long baseline of the SORCE mission allowed the analysis of estimated RV variation as a function of both time and wavelength. This data is presented in Figure \ref{fig:monthly} which shows the standard deviation of the estimated RV variation each month as a function of wavelength and time. Additionally, the TSI is over plotted on top of the TSI-estimated RV variation. The wavelength range that is considered unreliable is whited out along with months corresponding to instrument events (Table 1). Finally, we note that the SIM data vary in resolution with wavelength. The resolution functional form is given in Figure 10 of \citet{HAR05}. The resolution changes from 1nm at 300nm, to 10nm near 600nm, and to a maximum of $>$20nm near 1200nm.

Our results agree with our initial hypothesis that the NIR is reliably the region of lowest variability. Over most of the mission, the NIR shows a lower variation, although near the solar minimum (months 60-80) the visible reaches a variability on par with the NIR. The conclusion drawn from this graph is that, the NIR will show lower RV variability for solar type stars than the optical. If one has knowledge and is able to observe during the stellar cycle minimum, however, the visible variability should be on the same level as that in the NIR.

Additionally, we note that Fig.~\ref{fig:monthly} shows vertical streaks. Some of these coincide with known spectral lines, including lines sensitive to activity. The strongest line seen in this analysis (not shown) is actually in the UV and corresponds to the \ion{Mg}{2} H\&K lines. These lines are an order of magnitude more variable than any shown within Fig.~\ref{fig:monthly}. In the figure, some of the vertical streaks correspond to wavelengths of; $\sim$656 nm, around H$\alpha$; $\sim$590 nm, around the \ion{Na}{1} doublet; $\sim$519 nm, near the \ion{Mg}{1} triplet; and $\sim$392 and 396 nm, corresponding to Ca H \& K. We note that H$\alpha$ and Ca H \& K lines are known activity indicators \citep{HALPHA1,HALPHA2}. The \ion{Na}{1} D lines have also been explored as an activity tracer in cool stars \citep{ANDR97,DIAZ07}, but those lines are photon-dominated in hotter stars. All of these are approximate locations as the line width of the features in the plot is due to the resolution of the instrument which leads to uncertainty in the wavelength.

\subsection{Choice of $\kappa \delta V_c$}
In Section 5, we noted that our choice of $\kappa \delta V_c$ was taken from \citet{AIGRAIN12} and set to be 400 m s$^{-1}$. This value was fitted for by \citet{AIGRAIN12} from synthetic data which overlapped with our data set and was the value for the active Sun. We stress that while using a constant value prevents the values shown from being a perfectly accurate representation of reality, it does not prevent the comparison between wavelengths and with time. This value should decrease as the stellar cycle gets quieter, meaning that our values shown during the quiet period are upper limits. 

The choice of a constant $\kappa \delta V_c$ has an additional effect on the estimated RV noise as a function of wavelength. Different wavelengths probe different heights in the solar atmosphere and thus may be subject to different physical conditions, which can affect $\delta V_c$. Additionally, any wavelength dependance in the spot covering fraction would be due to a depth-dependent change in the spot size, which is not constrained by current data \citep{SPOTW}. This topic must be revisited for future F/F' applications as this is the first application to specific wavelengths instead of to large bandpasses or to the TSI.

\subsection{Observational Reality}
The overall reduction in F/F' estimated stellar RV noise in the NIR supports the possibility of the use of new NIR instruments, such as HPF, to observe bright solar-type stars. This claim is not without caveats. Specifically for F \& G stars, the NIR is hampered by the reduction in suitable spectral lines, an increase in telluric lines, and decreased emission from the stars \citep{BOTMR13}. Although the blackbody peaks of solar-types stars lie at visible wavelengths instead of in the NIR, they should still prove to be suitable targets if significantly brighter than the M \& K dwarfs that will be the primary targets of these instruments. The information content available and contamination from telluric lines are major obstacles that any NIR observers must consider. If the RV noise is indeed lower, as implied by our F/F' measurements, then these instruments may be able to unambiguously observe the RV signatures of smaller planets, in addition to supporting the observations of optical instruments. We feel that the NIR instruments will be especially useful in disentangling activity from real planetary signals for solar type stars.

\subsection{SIM Data Reduction}
The data used in this paper came from v19 of the SIM data release. Since this analysis was completed, v20 has been released using a new reduction algorithm. For consistency, we explored the v20 release. The results remain qualitatively the same, ie. variability is lower in the NIR, even though the two reductions show slightly different slopes and variability in SIM data.

\section{Summary}
We present the results of our F/F' analysis of the Sun using the SORCE spacecraft. Applying the method of \citet{AIGRAIN12} to the spectral irradiance data provided by the SIM instrument showed that the NIR has lower estimated RV noise than the optical. This is true over the entire stellar cycle suggesting that NIR instruments can play an important role in the detection and confirmation of low mass planets.

\subsection{Future Work}
The effects of stellar activity on the detection of exoplanets has been seen and continue to present a problem. Additional work should continue to determine both the best methods to confirm exoplanets (such as in what bandpass to observe) and on how to remove stellar RV signals from RV observations without affecting the signals of planets. Additional applications of the F/F' method to other stars as well as to the Sun can help refine the technique and prove its validity. Finally, a suitable NIR activity indicator should be sought to aid in observations of F, G, and K stars with NIR instruments.

\subsection{Acknowledgments}
This work was partially supported by funding from the Center for Exoplanets and Habitable Worlds. The Center for Exoplanets and Habitable Worlds is supported by the Pennsylvania State University, the Eberly College of Science, and the Pennsylvania Space Grant Consortium. This work was also supported directly by the Eberly College of Science at the Pennsylvania State University. We acknowledge support from NSF grant AST 1006676, AST 1126413, and AST 1310885 in our pursuit of precision radial velocities in the NIR. This work uses data from the SORCE/SIM team. We thank Steinn Sigurdsson for his input. Finally, we thank the anonymous reviewer for their time, effort, and suggestions. This work used graphics routines written by David Fanning. 


 
\end{document}